\documentstyle[12pt]{article}
\def\mybaselinestretch{1.0}   
\newif\ifdraftmode
\newif\ifdomarginnotes

\def\Titleofthispaper{%
                              Cluster Algorithm for Vertex Models
}%
   \def\Preprintnumber{\rule{0ex}{1ex}cond-mat/9211006 \\[.5ex]
                       FSU-SCRI-92-164}
   \def\Preprintdate  {November 1992}
\def\Authorsofthispaper{
     {\bf Hans Gerd Evertz$^{1}$, Gideon Lana$^{2}$},\\
     {\bf and Mihai Marcu$^{2}$}\\[3em]
     $^1\,$Supercomputer Computations Research Institute,        \\[-.2ex]
           Florida State University, Tallahassee, FL 32306    \\[-.2ex]
           evertz@scri.fsu.edu                                   \\[3ex]
     $^2\,$School of Physics and Astronomy,\\[-.2ex]
           Raymond and Beverly Sackler Faculty of Exact Sciences,\\[-.2ex]
           Tel Aviv University, 69978 Tel Aviv, Israel           \\[-.2ex]
           gidi@albert.tau.ac.il  \\[-.2ex]
           marcu@taunivm.bitnet      \\[3ex]
}%
\def\Abstracttext{We present a new type of cluster algorithm that
   strongly  reduces critical slowing down in simulations of
   vertex models. Since the clusters are closed paths of bonds,
   we call it the {\em loop algorithm}. The basic steps in
   constructing a cluster are the break-up and the freezing of
   vertices. We concentrate on the case of the F~model, which is
   a subset of the 6-vertex model exhibiting a Kosterlitz-Thouless
   transition. The loop algorithm is also applicable to
   simulations of other vertex models and
   of one and two-dimensional quantum spin systems.
\\[2ex] \noindent { PACS} numbers:
        02.70+d,  
        05.50+q,  
        75.10Jm,  
        68.35Rh   
}
    \newcommand{\note}[1]{}
    \ifdraftmode \ifdomarginnotes
                      \renewcommand{\note}[1]
                             {\marginpar{\raggedright\tiny #1}
                              \message{Note: #1}}
              \fi \fi
    \ifdraftmode  
             \fi
 \let\section=\subsection
 \let\subsection=\subsubsection
 \def\subsubsection#1{\subsection{#1}
      \par\note{CAUTION: subsection=subsubsection !}\par}
\newcount\hour   \newcount\minute
\hour  = \time   \divide\hour by 60
\minute= \hour   \multiply\minute by -60   \advance\minute by \time
\ifnum\minute<10 \def\Time{\number\hour:0\number\minute}
           \else \def\Time{\number\hour:\number\minute}  \fi

\newcommand{\figurebox}[2]{\fbox{\vbox to #1{\hbox to #2{\hfil}\vfil}}}
\newcommand{\str}{\rule{0ex}{2.7ex}}    
\newcommand{\tabhline}{\\[0.3ex] \hline \str}

\newcounter{itemnumber}

\let\mc=\multicolumn

\newcommand{\half}{{1\over2}}		    

\newcommand{\myeq}{\!=\!}

\newcommand{\mygt}{\!>\!}
\newcommand{\mygeq}{\!\geq\!}
\newcommand{\myleq}{\!\leq\!}

\def\3{\ss}

\def\kch{\half K_c}
\def\ropt{r_{\mbox{{\protect\scriptsize opt}}}}
\def\zcl{z^{\mbox{{\protect\scriptsize cl}}}}
\def\taucl{\tau^{\mbox{{\protect\scriptsize cl}}}}

\def\tauint{\tau_{\mbox{{\protect\scriptsize int}}}}

\def\zint{z_{\mbox{{\protect\scriptsize int}}}}
\def\eminusK{\mbox{e}^{-K}}
\def\eplusK{\mbox{e}^K}
\def\zatKc{0.71(5)}
\def\zatKch{0.19(2)}
\begin{document}
%
  \parbox[t]{20ex}{\ifdraftmode \fbox{\bf Draft version}   \fi}
  \hfill           \ifdraftmode \note{\today} \note{\Time} \fi
  \begin{tabular}[t]{l}
                     \rule{0ex}{1ex}\Preprintnumber \\[.5ex]
                     \rule{0ex}{1ex}\Preprintdate
  \end{tabular}
  \vfill
%
%
    \renewcommand{\thefootnote}{{\protect\fnsymbol{footnote}}}
         \vskip 2em
         \begin{center}
           {\Large \bf \vbox{\vspace{2ex}} \Titleofthispaper \par}\vskip1.5em
           {\normalsize \lineskip .5em
             \begin{tabular}[t]{c} \vbox{\vspace{3em}}  \Authorsofthispaper
             \end{tabular}\par}  \vskip 1em
         \end{center} \par 
    \renewcommand{\thefootnote}{{\protect\arabic{footnote}}}
\vfill
\renewcommand{\baselinestretch}{\mybaselinestretch}
\protect\small \protect\normalsize                  
\pagebreak[3]
\begin{abstract} {\protect\normalsize
                   \noindent \Abstracttext}
\end{abstract} \vfill
\thispagestyle{empty}
\setcounter{page}{0}
\newpage
\renewcommand{\baselinestretch}{\mybaselinestretch} 
\protect\small \protect\normalsize                  
%
%
%
 \section*{Introduction}
%
Cluster algorithms \cite{ClusterReviews,KandelDomany} are one of the few
known ways to overcome critical slowing down in Monte Carlo simulations.
Starting with \cite{SW} and continuing with new ideas like \cite{Wolff}
and \cite{FrustratedIsing}, most of the successful algorithms have dealt
with spin systems with two-spin interactions
(see however \cite{Z2gauge}). 

In {\em vertex models} \cite{Baxter,ReviewLieb} the dynamical variables
are localized on bonds, and the interaction is between all bonds meeting
at a vertex. Furthermore there are constraints on the possible bond
variable values around a vertex.

In this paper we present the {\em loop algorithm}, a new type of cluster
algorithm applicable to vertex models.
For usual spin systems most cluster algorithms start by ``freezing''
(also called ``activating'') or ``deleting'' bonds. Clusters are then
sets of sites connected by frozen bonds.
In the case of vertex models our idea is to define clusters as {\em
closed paths of bonds} (``loops''). To construct such clusters, we have
to perform operations at vertices that generalize the freeze-delete
procedure. In this context we introduce the concept of {\em break-up of
a vertex}.

For the sake of clarity we concentrate on the F~model, which is one of
the simplest vertex models.  We define it on an $L \times L$ square
lattice. Vertices are located at lattice sites. The bond variables take
the values $\pm 1$. They can be represented by arrows
(e.g.\ $+1$ means arrow up or right, $-1$ means arrow down or left).
At each vertex
we have the {\em constraint} that the number of incoming arrows equals
the number of outgoing arrows.  Thus there are six different vertex
configurations (six ``vertices''), as shown in fig.~1.
Their statistical weights $w(i)$, $i=1,\ldots,6$ are:
\begin{equation}\label{e1}
 w(i) = \left\{ \begin{array}{ll}
  \eminusK    &  i=1,2,3,4 \\
   1          &  i=5,6
 \end{array} \right. \; .
\end{equation}
The coupling $K\mygeq 0$ plays the role of inverse temperature. At $K_c
\myeq \ln 2$ there is a Kosterlitz-Thouless transition. The correlation
length is finite for $K \mygt K_c$ and  infinite for $K \myleq K_c$.

In what follows we start by presenting our new loop  algorithm. It turns
out that there is one free parameter in the algorithm. We discuss how to
choose an {\em optimal} value. Then we analyze the exponential
autocorrelation times at $K \myeq K_c$ and at $K \myeq K_c/2$. For the
optimum algorithm we find a dynamical critical exponent of $z(K_c) \myeq
\zatKc$ and $z(K_c/2) \myeq \zatKch$.
%
No critical slowing down is visible for the total energy.
%
We briefly show how to generalize our algorithm to more general six and
eight vertex models and how to use it for simulations of quantum spin
systems.
%
%
\begin{figure}[tb] \label{fig1}
\begin{center}
\setlength{\unitlength}{.0005\textwidth}
\begin{picture}(1700,350)
\thicklines
%
\newsavebox{\RIGHT}
\newsavebox{\LEFT}
\newsavebox{\UP}
\newsavebox{\DOWN}
\sbox{\RIGHT}{\put(  0,0  ){\vector( 1, 0){70}}\put(70,0 ){\line(1,0){30}}}
\sbox{\UP}   {\put(  0,0  ){\vector( 0, 1){70}}\put(0 ,70){\line(0,1){30}}}
\sbox{\LEFT} {\put(100,0  ){\vector(-1, 0){70}}\put(0 ,0 ){\line(1,0){30}}}
\sbox{\DOWN} {\put(  0,100){\vector( 0,-1){70}}\put(0 ,0 ){\line(0,1){30}}}
\newcommand{\VERTEX}[5]{\begin{picture}(0,0)
                              \put(-100,0   ){\usebox{#1}}       
                              \put(   0,0   ){\usebox{#2}}       
                              \put(   0,-100){\usebox{#3}}       
                              \put(   0,0   ){\usebox{#4}}       
                              \put(   0,-180){\makebox(0,0){#5}} 
                        \end{picture}}
%
\put( 100,200){\VERTEX{\RIGHT}{\RIGHT}{\UP}  {\UP}  {1}}
\put( 400,200){\VERTEX{\LEFT} {\LEFT} {\DOWN}{\DOWN}{2}}
\put( 700,200){\VERTEX{\RIGHT}{\RIGHT}{\DOWN}{\DOWN}{3}}
\put(1000,200){\VERTEX{\LEFT} {\LEFT} {\UP}  {\UP}  {4}}
\put(1300,200){\VERTEX{\RIGHT}{\LEFT} {\DOWN}{\UP}  {5}}
\put(1600,200){\VERTEX{\LEFT} {\RIGHT}{\UP}  {\DOWN}{6}}
\end{picture}
\caption[fig1]{\parbox[t]{.8\textwidth}{
               The six vertex configurations.
               The labels $1,\ldots,6$
                follow standard conventions \cite{Baxter}. }}
\end{center}
\end{figure}
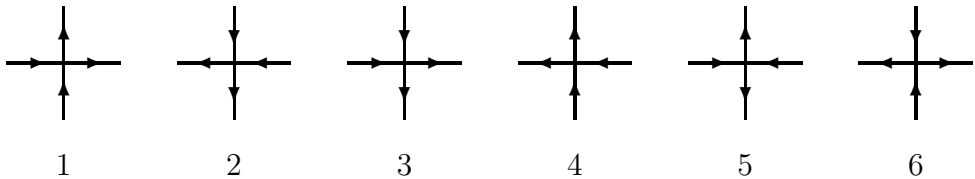
%
%
 \section*{The Loop Algorithm}
%
If we regard the arrows on bonds as a vector field, the constraint at
the vertices is a zero-divergence condition. Therefore every
configuration change can be obtained as a sequence of {\em loop-flips}.
By ``loop'' we denote an oriented, closed,
non-branching (but possibly self-intersecting)
path of bonds, such that all arrows along the path point in the
direction of the path. A loop-flip reverses the direction  of
all arrows along the loop.

Our cluster algorithm performs precisely such operations, with
appropriate probabilities.
It constructs closed paths consisting of one or
several loops without common bonds. All loops in this path are flipped
together.

We shall construct the path iteratively, following the direction of the
arrows. Let bond $b$ be the latest addition to the path. The arrow on
$b$ points to a new vertex $v$. There are two outgoing arrows at $v$,
and what we need is a unique prescription for continuing the path
through $v$. This is provided by a {\em break-up} of the vertex $v$.
In addition to the break-up, we have to allow for {\em freezing} of $v$.
By choosing suitable probabilities for break-up and freezing we shall
satisfy detailed balance.

The {\em break-up} operation is defined by splitting $v$ into two
corners, as shown in fig.~2.
At any corner one of the arrows points towards $v$, while the
other one points away from $v$. Thus we will not allow e.g.\ the ul--lr
break-up for a vertex in the configuration 3.
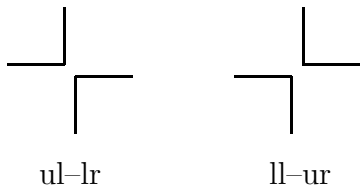
\begin{figure}[tb] \label{fig2}
\begin{center}
\setlength{\unitlength}{.0005\textwidth}
\begin{picture}(620,300)
\thicklines
\put(100,210){\line(-1, 0){100}}
\put(100,210){\line( 0, 1){100}}
\put(120,190){\line( 1, 0){100}}
\put(120,190){\line( 0,-1){100}}
\put(110,20){\makebox(0,0){ul--lr}}

\put(500,190){\line(-1, 0){100}}
\put(500,190){\line( 0,-1){100}}
\put(520,210){\line( 1, 0){100}}
\put(520,210){\line( 0, 1){100}}
\put(515,20){\makebox(0,0){ll--ur}}
%
%
%
\end{picture}
\caption[fig2]{ \parbox[t]{.80\textwidth}{
                The two break-ups of a vertex:
                ul--lr (upper-left--lower-right) and
                ll--ur (lower-left--upper-right)
              } }
\end{center}
\end{figure}
%
%
A ``corner flip'' is a flip of both arrows. For a given break-up, we
only allow the configuration changes resulting from independent corner
flips. This preserves the zero divergence condition at $v$.
Notice that a single corner flip transforms a vertex of weight
$1$ into a vertex of weight $\eminusK$ and vice-versa.
Detailed balance is satisfied with the following probabilities for
choosing a given break-up:
\begin{equation}\label{e2}
P_{\,\mbox{ul-lr}}(i) = \left\{ \begin{array}{ll}
  r \eplusK                 & i=1,2 \\
  0                         & i=3,4 \\
  r                         & i=5,6
 \end{array} \right. \; , \quad \quad
P_{\,\mbox{ll-ur}}(i) = \left\{ \begin{array}{ll}
  0                         & i=1,2 \\
  r \eplusK                 & i=3,4 \\
  r                         & i=5,6
 \end{array} \right. \; ;
\end{equation}
$r$ is a free parameter for now.

{\em Freezing} of a vertex means that its weight must not change. Since
there are only two different vertex weights, we introduce two freezing
probabilities. They are already determined by the requirement that for a
given vertex configuration the sum of freezing and break-up
probabilities must be one:
\begin{equation}\label{e3}
P_{\mbox{freeze}}(i) = \left\{ \begin{array}{ll}
  1-r\eplusK     &  i=1,2,3,4 \\
  1-2r           &  i=5,6
 \end{array} \right. \; .
\end{equation}
The range of possible values for $r$ is now obtained by requiring
that all probabilities are between zero and one:
\begin{equation}\label{e3a}
 0 \leq r \leq \min(\half,\eminusK) \; .
\end{equation}
%

Assume now that we have broken or frozen all vertices. Starting from a
bond $b_0$, we proceed to construct a closed path by moving in the arrow
direction. As we move from vertex to vertex, we always have a unique way
to continue the path. If a vertex is broken, we enter and leave  it
along the same corner. If the vertex is frozen and of type 1, 2, 3 or 4,
we pass through it on a straight line. At such vertices the path may be
self-intersecting.
Finally, if the latest bond $b$ added to the cluster points to a
frozen vertex $v$ of type 5 or 6, the path continues both to the right
and to the left of $b$,
i.e. we start a new loop at $v$.
The two loops have to be flipped together.
In general, the zero-divergence condition
guarantees that all loops will eventually close.

The break-or-freeze decision for all vertices determines a unique
partitioning of the lattice into closed paths that can be flipped
independently. We choose to perform single cluster updates, i.e.\ we
``grow'' a {\em single path} from a random starting bond $b_0$, and flip
it. The break-or-freeze decision is only needed for the vertices along
the path. Thus the computer time for one path is proportional to the
length of that path.

It is easy to see that our algorithm is correct.
The proof of detailed balance is completely analogous to that for other
cluster algorithms \cite{ClusterReviews,KandelDomany}. The main
ingredient here is that $P_{\,\mbox{ul-lr}}$ and $P_{\,\mbox{ll-ur}}$
already satisfy detailed balance locally.
Furthermore, it is not difficult to see that any two allowed
configurations can be connected by a finite number of cluster flips.
Thus a finite power of the Markov matrix is ergodic.

How do we choose an optimal value for the parameter $r$ ?
We have seen that freezing of a vertex of type 5 or 6 forces us to
flip two loops together.
If we had broken it up instead, we might have been allowed to
flip the two loops independently. Thus more freezing leads to larger clusters.
We conjecture that the {\em least possible freezing is optimal}.
This is confirmed by numerical tests (see below).
{}From eq.\ (\ref{e3}) we then obtain
\begin{equation}\label{e4}
 \ropt = \left\{ \begin{array}{ll}
                    \half     & \quad K \myleq K_c   \\[.5ex]
                    \eminusK  & \quad K \mygeq K_c
                 \end{array} \right. \; .
\end{equation}
By maximizing $r$ we also minimize the freezing probability
for vertices of type 1, 2, 3 and 4.
Notice that if we choose $r=\ropt$,
then for $K \myleq K_c$  vertices of type 5 and 6 are never frozen,
so every path consists of a single loop.
For $K \mygt K_c$ on the other hand,
vertices of type 1, 2, 3 and 4 are never frozen, so we do not
continue a path along a straight line through any vertex.

There are some distinct differences between our loop-clusters and more
conventional spin-clusters. For spin-clusters, the elementary objects
that can be flipped are spins; freezing binds them together into
clusters. Our closed loops on the other hand may be viewed as a part of
the {\em boundary} of spin-clusters
(notice that the boundary of spin clusters may
contain loops inside loops). It is reasonable to expect that in
typical cases, building a loop-cluster will cost less work than for a
spin-cluster. This is an intrinsic advantage of the loop algorithm.

This can be exemplified nicely for the F~model, where a spin-cluster
algorithm -- the VMR algorithm \cite{BCSOScluster} -- is also available.
At $K_c$ one can see that if we use $r=\ropt$, loop-clusters are indeed
parts of the boundary of VMR spin-clusters. Since flipping a
loop-cluster is not the same as flipping a VMR cluster, we expect the
two algorithms to have different performance. We found (see
\cite{BCSOScluster} and the next section) that in
units of clusters, the VMR algorithm is more efficient, but
in work units, which are basically units of CPU time, the loop algorithm
wins. At $K_c/2$, where the loop-clusters are not related \cite{LargePaper}
to the boundary of VMR clusters,
we found the loop algorithm to be more efficient both in units of
clusters and in work units, with a larger advantage in the latter.

 \section*{Performance}
%
We tested our new algorithm on $L \times L$ square lattices
with periodic boundary conditions,
both at the transition point $K_c$ and
at $\kch$ deep inside the massless phase.
%
%
We carefully analyzed autocorrelation functions
and determined the exponential
autocorrelation time $\tau$.
At infinite correlation length,
{\em critical slowing down} is quantified by the relation
\cite{ClusterReviews}
\begin{equation}\label{csd}
\tau \propto L^{\mbox{\protect\small $z$}} \, .
\end{equation}
Local algorithms are slow, with $z \approx 2$.
For comparison, we performed runs with a local algorithm
that flips arrows around elementary plaquettes with
Metropolis probability, and indeed found $z = 2.2(2)$ at $K=K_c$.

In order to make sure that we do observe the slowest mode of the Markov
matrix we measured a range of quantities and checked that they exhibit
the same $\tau$.
As in \cite{BCSOScluster}, the slowest mode is strongly coupled to
the sublattice energy%
{}.
%
The two sublattice energies%
\cite{BCSOScluster}
 add up to the total energy. The constraints of the model
cause them to be strongly anticorrelated.
Within our precision the
true value of $\tau$ is {\em not} visible from autocorrelations of the
total energy, which decay very quickly. Only for the largest lattices do
we see a small hint of a long tail in the autocorrelations.
A similar situation occurred in \cite{BCSOScluster}, where,
when decreasing the statistical errors,
the decay governed by the true $\tau$ eventually became visible.
Note that as a consequence of this situation,
the so-called ``integrated  autocorrelation time''
\cite{ClusterReviews} is much smaller than $\tau$, and it would be
completely misleading to evaluate the algorithm based only on its values.
%

We shall quote autocorrelation times $\tau$ in units of
``sweeps'' \cite{ClusterReviews}.
We define a sweep such that on average each bond
is updated once during a sweep.
Thus, if $\taucl$ is the autocorrelation time in units of clusters%
, then
$\tau = \taucl \times
       <\mbox{cluster size}> / {(2L^2)}$.
Each of our runs consisted of between 50000 and 200000 sweeps.
Let us also define $\zcl$ by
$\taucl \propto L^{\mbox{\protect\small $\zcl$}}$,
and a cluster size exponent $c$ by
$<\mbox{cluster size}> \, \propto L^{\mbox{\protect\small $c$}}$.
We then have:
\begin{equation}\label{taucl}
   z = \zcl - (2-c) \, .
\end{equation}

\begin{table}[tb]
 \centering
\vskip2ex
\begin{tabular}{|r|r|r|}
\hline\str
 $L$ & \mc{1}{c|}{$K=K_c$} & \mc{1}{c|}{$K=\kch$}
\\[.5ex]\hline \str
  8 & 1.8(1)   & 4.9(4) \\
 16 & 3.0(2)   & 5.6(2) \\
 32 & 4.9(4)   & 6.2(3) \\
 64 & 7.2(7)   & 7.4(3) \\
128 &15.5(1.5) & 8.3(2) \\
256 &20.5(2.0) &
\tabhline
 $z$  & \zatKc  &\zatKch
\\[.3ex] \hline
\end{tabular}
 \caption[dummy]{\label{tab1} \parbox[t]{.85\textwidth}{
                              Exponential autocorrelation time $\tau$
                              at $r \!=\! \ropt$,
                              and the resulting \mbox{dynamical}
                              critical exponent $z$.
                          }}
\end{table}

Table 1 shows the autocorrelation time $\tau$ for the optimal choice
$r\myeq\ropt$.
At $K=\kch$, deep inside the massless phase, critical slowing down is almost
completely absent. A fit according to eq.\ \ref{csd} gives $z=\zatKch$.
The data are also consistent with $z=0$ and only logarithmic growth.
For the cluster size exponent $c$
we obtained $c=1.446(2)$, which points to
the clusters being quite fractal.
At the phase transition $K = K_c$ we obtained
$z=\zatKc$, which is still small.
The clusters seem to be less fractal: $c=1.060(2)$.

We noted above that at
$K=K_c$ and for the optimal choice of $r$, the loop-clusters are related
to the VMR spin-clusters. In \cite{BCSOScluster} we obtained for
the VMR algorithm at $K=K_c$ the result $\zcl=1.22(2)$, but we
had $c=1.985(4)$, which left us with $z=1.20(2)$. In this case
it is the smaller dimensionality of the clusters that make the
loop algorithm more efficient.

As mentioned, no critical slowing down is visible for the integrated
autocorrelation time of the total energy.
At $K=K_c$, $\tauint(E)$ is only 0.80(2) on the largest lattice,
and we find $\zint(E) \approx 0.20(2)$.
At $K=\kch$, $\tauint(E)$ is $1.1(1)$ on all lattice sizes,
so $\zint(E)$ is zero.

What happens for non-optimal values of $r\,$?
Table \ref{tab2} shows our results on the dependence of $z$ on $r$.
$z$ rapidly increases as $r$ moves away from $\ropt$.
This effect seems to be stronger at $\kch$ than at $K_c$.
We thus see that the optimal value of
$r$ indeed produces the best results, as conjectured from
our principle of {\em least possible freezing}.

%
\begin{table}[tb]  \centering
\begin{tabular}{|c|c|r|}
\hline\str
 $K$   & $r$ & \mc{1}{c|}{$z$}
        \\[.5ex] \hline  \str
$\kch$ & 0.500 &$ \zatKch$    \\[.3ex]
$\kch$ & 0.450 &$ 1.90(5)$    \\[.3ex]
$\kch$ & 0.400 &$\geq2.6(4)$
\tabhline
$K_c$  & 0.500 &$ \zatKc $    \\[.3ex]
$K_c$  & 0.475 &$ 0.77(6)$    \\[.3ex]
$K_c$  & 0.450 &$ 0.99(6)$    \\[.3ex]
$K_c$  & 0.400 &$\geq2.2(1)$
\\[.3ex] \hline
\end{tabular}
 \caption[dummy]{\label{tab2}\parbox[t]{.8\textwidth}{
       Dependence of the dynamical critical exponent
       $z$ on the parameter $r$.
       We use ``$\geq$'' where for our lattice sizes $\tau$
       increases faster than a power of $L$.
   }}
\end{table}

In the massive phase close to $K_c$,
we expect \cite{LargePaper} that
$z(K_c)$ will determine the behaviour of $\tau$
in a similar way as in ref.\ \cite{BCSOScluster}.
To confirm this, a finite size scaling analysis of
$\tau$ is required.

 \section*{Generalizations and Outlook}
%
For the sake of clarity we have described our approach
in terms of the F~model only.
It has however a much wider range of applicability.
We will give a detailed description elsewhere \cite{LargePaper}.
Here we shall  mention only a few highlights.

Our ``break-up of vertices'' and subsequent path flip
{\em automatically} satisfies the constraints of the F~model.
General six {\em and} eight vertex models \cite{Baxter}
with arrow flip symmetry have related constraints.
By using the framework of Kandel and Domany \cite{KandelDomany}
and the principle of minimal freezing,
we can generalize the break-up operation \cite{LargePaper} to obtain
efficient algorithms for these cases too.
Algorithms for more general vertex models can be engineered
along the same lines.

Particularly promising is the possibility of
{\em accelerating Quantum Monte Carlo simulations} \cite{QMC,LargePaper}.
Quantum spin systems in one and two dimensions can be mapped into
vertex models in $1+1$ and $2+1$ dimensions via the Trotter formula
and suitable splittings of the Hamiltonian \cite{QMC}.
The simplest example is the spin $\half$ $xxz$ quantum chain, which is
mapped into the 6-vertex model.
For higher spins, more complicated
vertex models result (e.g.\ 19-vertex model for spin one).

For $(2+1)$ dimensions, different splittings of the Hamiltonian can
lead to geometrically quite different situations
\cite{LargePaper,QMC}.
We can e.g.\ choose between 6-vertex models on a complicated $2+1$
dimensional lattice, and models on a {\em bcc} lattice with 8
bonds (and a large number of configurations) per vertex.
In all these cases, the constraints are of a similar nature
as in the F~model,
and our approach of constructing and updating clusters which are
{\em paths} can be applied in a straightforward way.
%

Notice also that in our approach it is easy to change global properties
like the number of world lines or the winding number (see \cite{QMC}).

Recently we received a paper  by Wiese and Ying \cite{WieseYing}
on a different cluster algorithm for spin $\half$ quantum spin systems.
After mapping to a vertex model
similar to the one we refer to,
they combine vertices into ``block-spins''  which are then used
in a standard spin-cluster construction. This approach
restricts the possible updates of the arrows.
In our language, their clusters are sets of loops that are frozen together,
i.e.\ that have to be flipped together.
For some interesting cases, e.g.\ the one-dimensional
Heisenberg ferromagnet
and two-dimensional Heisenberg ferromagnet and anti-ferromagnet,
the additional freezing leads to the problem of
{\em frustration} for the block-spin clusters.
We expect our algorithm to perform
better in these cases, both because our clusters have less loops and
because of the added flexibility offered by the
possibility to optimize.
%

\section*{Conclusions}
%
We have presented a new type of cluster algorithm.
It flips closed paths of bonds in vertex models.
Constraints are automatically satisfied.
We have succesfully tested our algorithm for the F~model and found
remarkably small dynamical critical exponents.

There are many promising and
straightforward applications of our approach,
to other vertex models, and to 1+1 and 2+1 dimensional
quantum spin systems.
Investigations of such systems are in progress.

\vskip3ex
\section*{Acknowledgements}
This work was supported in part by the German-Israeli
Foundation for Research and Development (GIF) and
by the Basic Research Foundation of
the Israel Academy of Sciences and Humanities.
We used about 50 hours of Cray {\sc YMP} processor time.
We would like to express our gratitude
to the HLRZ at KFA J\"ulich 
and to the DOE for providing the necessary computer time.

   \pagebreak

\end{document}